\documentclass[aps,prl,twocolumn,groupedaddress,showpacs,showkeys]{revtex4-1}

\usepackage{graphicx}
\usepackage{array}
\usepackage{SIunits}
\usepackage{color} 
\begin{document}


\title{Efficient quantum memory using a weakly absorbing sample}


\author{Mahmood Sabooni, Qian Li, Stefan Kr\"{o}ll, Lars Rippe}
\affiliation{Department of Physics, Lund University, P.O.~Box 118, SE-22100 Lund, Sweden}



\begin{abstract}
A light-storage experiment with a total (storage and retrieval) efficiency $\eta=58 \pm 5\%$ is carried out by enclosing a sample, with a single pass absorption of $10\,\%$, in an impedance-matched cavity. The experiment is carried out using the Atomic Frequency Comb (AFC) technique in a praseodymium-doped crystal ($0.05\,\%\, Pr^{3+}\!:\!Y_2SiO_5$) and the cavity is created by reflection coating the crystal surfaces. The AFC technique has previously by far demonstrated the highest multi-mode capacity of all quantum memory concepts tested experimentally. We claim that the present work shows that it is realistic to create efficient, on-demand, long storage time AFC memories.
\end{abstract}


\maketitle

\par
A quantum memory which has the ability to map onto, store in, and later retrieve the quantum state of light from matter is an important building block in quantum information processing \cite{Hammerer2010}. Quantum memories are expected to become vital elements for long distance quantum key distribution \cite{Duan2001,Sangouard2008}. Quantum computing based on linear optics schemes \cite{Kok2007}, signal synchronization in optical quantum processing \cite{Knill2001,Nunn2012}, the implementation of a deterministic single-photon source \cite{Chen2006}, and precision measurements based on mapping of the quantum properties of an optical state to an atomic ensembles \cite{Lvovsky2009} are other applications of quantum memories. For most of the applications mentioned, high performance will be required in terms of high efficiency \cite{Hosseini2011,Hedges2010}, on-demand readout, long storage time \cite{Longdell2005,Heinze2012}, multi-mode storage capacity \cite{Bonarota2011,Usmani2010} and broad bandwidth \cite{Saglamyurek2011}.

\par
Many protocols have been proposed to realize an efficient quantum memory, these include electromagnetically induced transparency (EIT) \cite{Fleischhauer2000a}, off-resonant Raman interactions \cite{Reim2010}, controlled reversible inhomogeneous broadening (CRIB) \cite{Moiseev2001,Nilsson2005,Kraus2006}, gradient echo memory (GEM) \cite{Hetet2008a}, and atomic frequency combs (AFC) \cite{Afzelius2009}. The most impressive storage and retrieval efficiencies so far, 87\% \cite{Hosseini2011} and 69\% \cite{Hedges2010}, were achieved in hot atomic vapor and rare-earth doped crystals, respectively using the GEM technique. Additionally, 43\% storage and retrieval efficiency using EIT in a hot Rb vapor \cite{Phillips2008} and 35\% using AFC in a rare-earth doped crystal \cite{Amari2010} were attained.

\par
\begin{figure}[ht]
    \includegraphics[width=8cm]{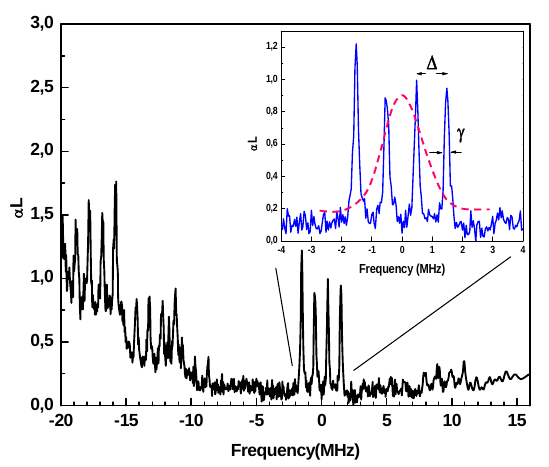}
    \caption{(Color online) The absorption profile created through the optical pumping process is shown. This measurement is done in a small part of the cavity crystal which has been left un-coated. From the $Pr$ inhomogeneous absorption profile, four AFC peaks, all absorbing on the $|\pm\frac{5}{2}g\rangle \mapsto |\pm\frac{5}{2}e\rangle$ transition are created in the interval $-1.5\,M\!H\!z$ to $1.5\,M\!H\!z$. The spectral content of the storage pulse is shown schematically in the inset by the red dashed line across the AFC structure which has peak width $\gamma$ and peak separation $\Delta$. The finesse of the AFC structure is $F_{AFC}=\frac{\Delta}{\gamma}$.}
    \label{Readout}
\end{figure}

The AFC technique \cite{Afzelius2009} is employed in this paper since the number of (temporal) modes which can be stored in a sample is independent of the optical depth ($d$) of the storage material, in contrast to other quantum memory approaches. For a comparison, the number of modes which can be stored is proportional to $\sqrt{d}$ in EIT and off-resonant Raman interactions and to $d$ in CRIB \cite{Afzelius2009}, which means the optical depth needs to be increased for these techniques to be able to store many modes. An AFC structure consists of a set of (artificially created) narrow absorbing peaks with equidistant frequency spacing $\Delta$ and uniform peak width $\gamma$ (see the inset in Fig. \ref{Readout}). An input (storage) field (at time $t=0$) which spectrally overlaps the AFC structure will be absorbed and leave the absorbers (in our case the $Pr$ ions) in a superposition state \cite{Afzelius2009}. If the coherence time is long compare to $1/\Delta$, a collective emission due to constructive interference ( similar as for the modes in a mode-locked laser) will occur at time $t_{echo}=1/\Delta$. On-demand retrieval of the stored input field is possible by transferring the ground-excited state superposition to a spin-level superposition as demonstrated in Ref. \cite{Afzelius2010}.

\par
High storage and retrieval efficiency is one of the main targets of quantum memories and this relies on strong coupling between light and matter \cite{Hammerer2010}. One approach for studying light-matter interaction is based on the high finesse cavity-enhanced interaction of light with a single atom \cite{Raimond2001}. Another alternative for increasing the coupling efficiency of a quantum interface between light and matter is using an optically thick free space atomic ensemble \cite{Hammerer2010}. In this paper, we combine the advantages of both approaches to implement an efficient quantum interface in a weakly absorbing solid state medium. Within the ensemble approach several experimental realization from room-temperature alkali gases \cite{Julsgaard2004}, to alkali atoms cooled and trapped at temperature of a few tens or hundreds of microkelvin \cite{Simon2007a} have been investigated. Among the ensemble-based approaches impurity centers in a solid state crystal is a powerful alternative for quantum memories because of the absence of atomic movement.

\begin{figure}[ht]
    \includegraphics[width=8cm]{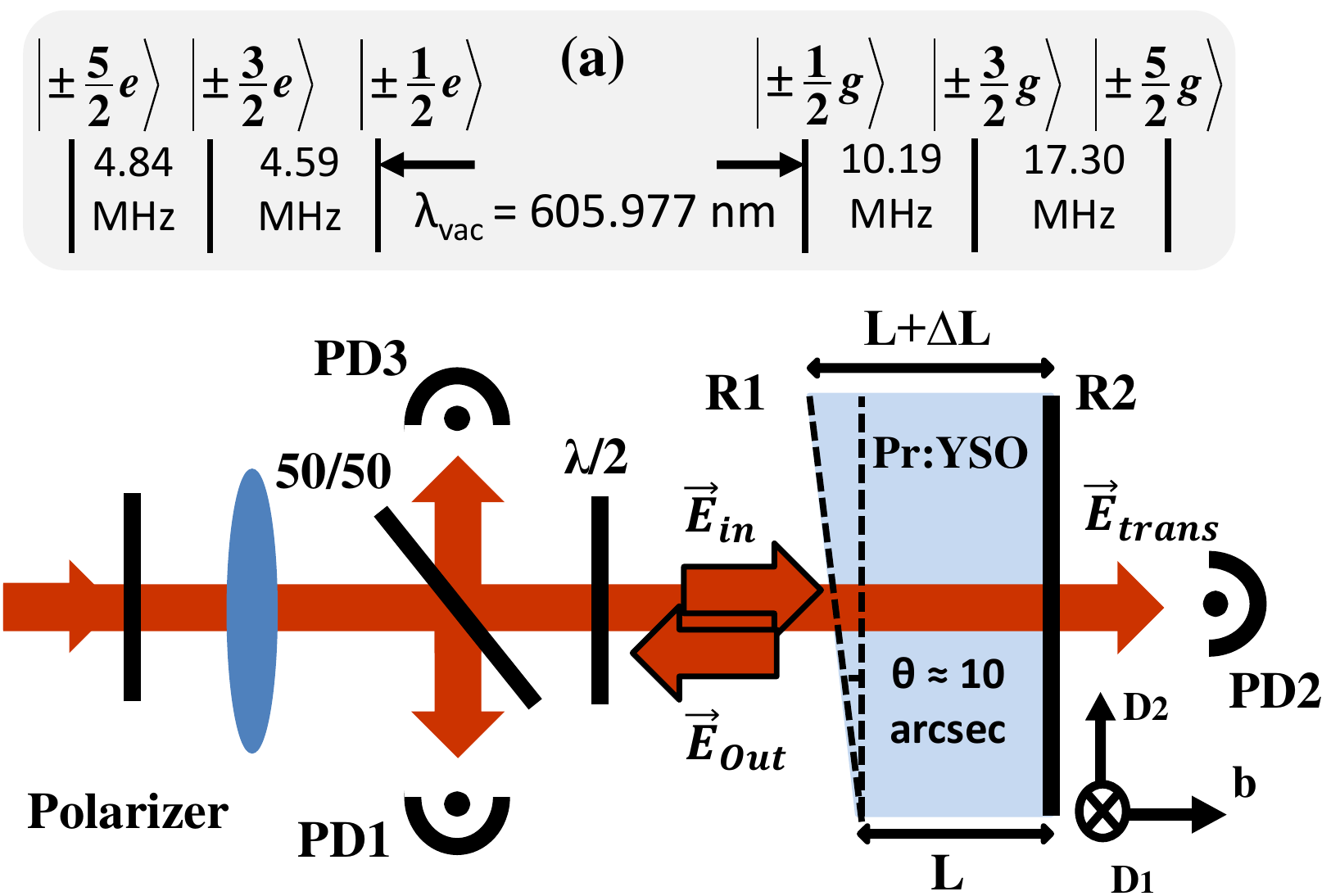}
    \caption{(Color online) Part of the experimental set up. A frequency stabilized ($<1\,k\!H\!z$ line-width) dye laser at $\lambda_{vac}=605.977\,nm$ is employed as light source. A 50/50 beam splitter splits off part of the input light onto a photo diode (PD1), which is used as an input light intensity reference. Two other photo diodes monitor the transmitted (PD2) and the reflected (PD3) light from the cavity. The cavity length along the $b$ axis is $L=2\,mm$ and the crystal diameter in the $(D_1,D_2)$ plane is $12\;mm$. b, $D_1$ and $D_2$ are principal axes of the crystal \cite{GuokuiLiu2005}. A beam waist of about $100\;\mu m$ is created at the crystal center via a lens with $f=400\:mm$. A half-wave plate ($\lambda/2$) is employed to align the polarization direction to a principal axis of the cavity crystal. The input and output facet of the crystal has $R1=80\%$ and $R2=99.7\%$ reflectivity. A small part of the cavity crystal is left un-coated for measurements without a cavity effect. (a) The hyperfine splitting of the ground $|g\rangle$ and excited $|e\rangle$ state of the $^3H_4-\!^1D_2$ transition of site I in $Pr^{3+}\!:\!Y_2SiO_5$ \cite{Rippe2005}.}
    \label{setup}
\end{figure}

\par
The objective of this paper is to demonstrate a quantum memory with high storage and retrieval efficiency, with the added benefit of being achievable in a weakly absorbing medium. Another benefit is the short crystal length ($2\;mm$), and small physical storage volume ($\ll\,mm^3$). This can simplify the implementation of long term storage in the hyperfine levels, as will be further discussed at the end of the paper. Our work is based on the proposals in Refs. \cite{Afzelius2010a,Moiseev2010}, where it is shown that close to unity storage and retrieval efficiency can be obtained, using an atomic ensemble in an impedance-matched cavity. A cavity can be impedance-matched, by having the absorption per cavity round trip, ($1- e^{-2\alpha L}$), equal to the transmission of the input coupling mirror, ($1-R1$), while the back mirror is $100\%$ reflecting, $R1=e^{-2\alpha L}$, where $\alpha$ is the absorption coefficient and $L$ is the length of the optical memory material. For this impedance-matching condition, all light sent to the cavity will be absorbed in the absorbing sample inside the cavity and no light is reflected.

\par
The storage cavity is made up of a $2\,mm$ long $0.05\%\;Pr^{3+}\!:\!Y_2SiO_5$ crystal, see Fig. \ref{setup}. To reduce the complexity of the alignment and reduce losses, the crystal surfaces are reflection coated directly, rather than using separate mirrors. The two cavity crystal surfaces are not exactly parallel as shown in Fig. \ref{setup} ($\theta\approx10\;arcsec$). Part of the incoming field $\overrightarrow{E}_{in}$ will be reflected ($\overrightarrow{E}_{refl}$) at the first mirror surface ($R1$), see Fig. \ref{setup}. The field leaking out through $R1$ from the cavity, $\overrightarrow{E}_{leak}$, is coherently added to  $\overrightarrow{E}_{refl}$ such that $\overrightarrow{E}_{out}=\overrightarrow{E}_{refl}+\overrightarrow{E}_{leak}$. At the impedance-matched condition, $\overrightarrow{E}_{refl}$ and $\overrightarrow{E}_{leak}$ differ in phase by $\pi$ and have the same amplitude $\mid\overrightarrow{E}_{ref}\mid=\mid\overrightarrow{E}_{leak}\mid$. This means that the light intensity at the reflection detector (PD3) should ideally vanishes if this condition satisfied. 

\begin{figure}[ht]
    \includegraphics[width=8cm]{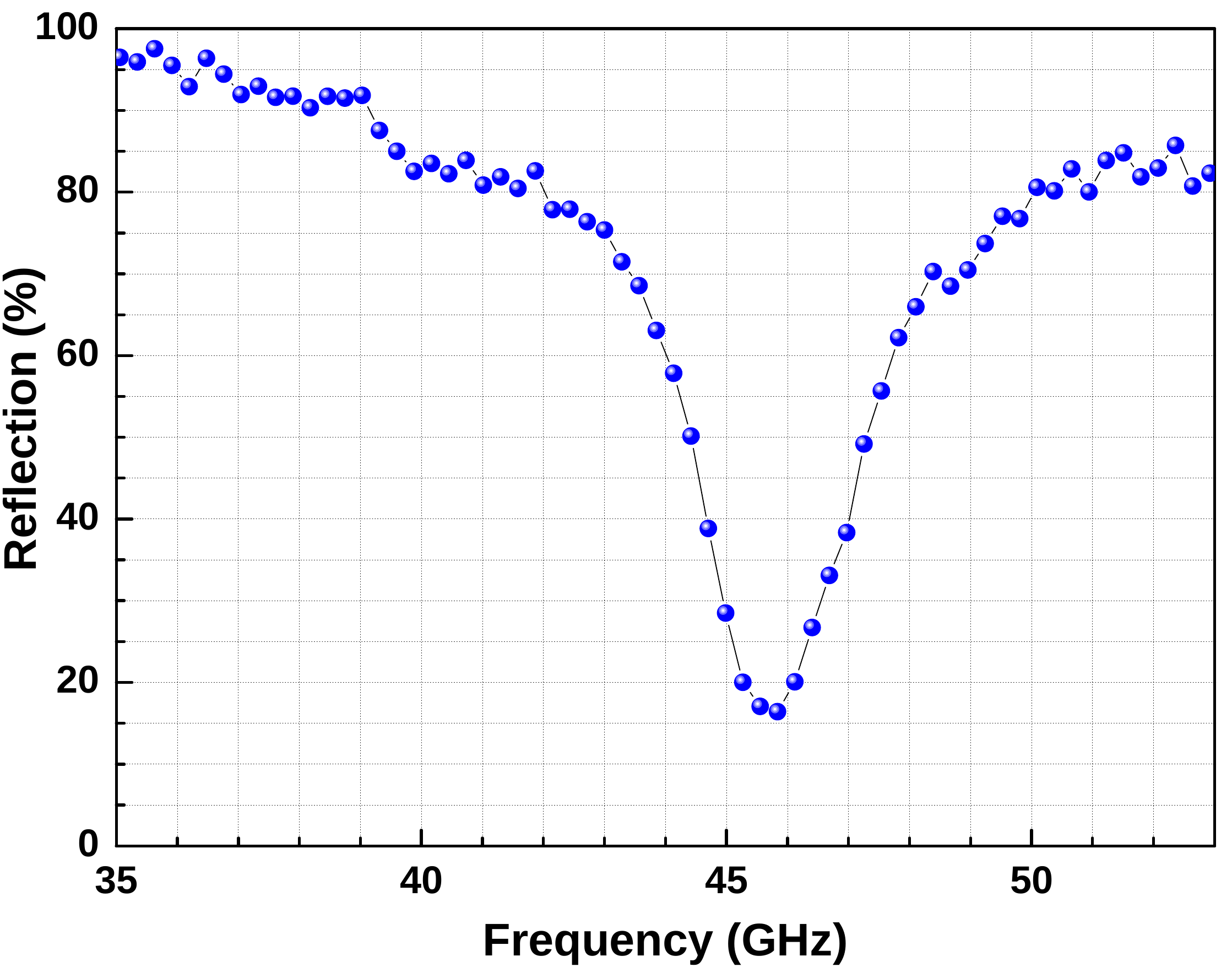}
    \caption{(Color online) The normalized reflected light at PD3 is plotted against the frequency offset from the $Pr^{3+}\!:\!Y_2SiO_5$ inhomogeneous line center. The crystal was translated perpendicular to the input beam (see Fig. \ref{setup}) and the graph is a frequency scan for the position that gave the best impedance matching. At the impedance-matching condition the reflected light detected at PD3 should vanish. The best impedance-matched condition without memory preparation (i.e. spectral manipulation by the absorption profile, see text) was a reflection of $16\%$ ($84\%$ absorption) which was measured about $45\;GHz$ above the inhomogeneous broadening center frequency which is at zero in this figure.
    \label{imp_match}}
\end{figure}

\par
To optimise the impedance-matching in the cavity crystal without memory preparation, firstly, the angle between the input beam and the beam reflected from the cavity is exactly overlapped within an angle smaller than a few $arcmin$. Secondly, due to a small cavity surface wedge ($\theta\approx10\;arcsec$) the effective cavity length can be optimized by translating the cavity perpendicular to the beam propagation direction. Here a sub $\mu$m accuracy (Attocube system, ANCz150) translation stage was used. Thirdly, a weak Gaussian pulse ($\tau_{FWHM}=250\;ns$) with a small pulse area and repetition rate of $20\:H\!z$ was injected into the cavity while the laser frequency was slowly scanned during $60\;s$ across the inhomogeneous $Pr$ ion absorption line. The cavity crystal was moved spatially to find the best impedance-matched point and for each point a $18\;G\!H\!z$ laser scan was carried out. To calibrate our measurement against laser amplitude variations, the integrated pulse intensity of the light detected at PD3 is normalized against the integrated intensity of the pulse detected at PD1 in Fig. \ref{setup}. The background light at PD3 is negligible. The calibrated reflected part of the input pulse is plotted versus the frequency offset from the $Pr^{3+}\!:\!Y_2SiO_5$ inhomogeneous line center in Fig. \ref{imp_match}. This measurement shows a maximum of about $84\;\%$ of the input energy was absorbed about $45\;GHz$ above the inhomogeneous broadening center frequency. This will set an upper limit for the achievable storage and retrieval efficiency in the present setup. The inhomogeneous absorption line-width in this crystal is about $10\,G\!H\!z$. Therefore, assuming a Lorentzian line-shape, the absorption will be about $1\,\%$ of the line center absorption where the impedance-matched condition is fulfilled. Due to the absorption tailoring during the memory preparation (to be discussed below), the impedance-matching condition will be fulfilled closer to the the inhomogeneous line center in the actual storage experiment, but the measurement establish the losses caused by spatial mode mismatching.

\par
To demonstrate a quantum memory based on the AFC protocol, firstly, a transparent (non-absorbing) spectral transmission window within the $Pr$ ion absorption profile was created using optical pumping. An accurate description of the pulse sequences required for creating such a transparency window, which henceforth is called a spectral pit, is given in Ref. \cite{Amari2010}. Due to the strong dispersion created by the spectral pit \cite{Walther2009a}, the cavity free spectral range (FSR) and the cavity line-width are reduced by 3-4 orders of magnitude. The reduction can be understand as follows. The cavity FSR is $\Delta\nu_{mode}=\frac{c_0}{2L}\frac{1}{n_g}$ \cite{Siegman1985} where $c_0$ is speed of light in the vacuum, $L$ is the cavity length, and $n_g$ is the group refractive index. $n_g=n_r(\nu)+\nu\frac{dn_r(\nu)}{d\nu}$ and $n_r(\nu)$ is the real refractive index as a function of frequency, $\nu$. In case of no or constant absorption, the dispersion term is negligible compare to the real refractive index ($n_r(\nu)\gg\nu\frac{dn_r(\nu)}{d\nu}$). The FSR of this cavity with no absorbing material is about $40\:GHz$. In the presence of sharp transmission structures ($n_r(\nu)\ll\nu\frac{dn_r(\nu)}{d\nu}$) a dramatic reduction of the cavity FSR and the cavity line-width can occur. In our case $\nu\frac{dn_r(\nu)}{d\nu}>1000n_r(\nu)$ and the reduction is $>$ 3 orders of magnitude. Because of this strong FSR reduction, there will be at least one cavity resonance peak within the spectral pit prepared for the AFC structure. A more detailed description of the cavity FSR reduction is given in Ref. \cite{Sabooni2012a}. Translating the crystal perpendicular to the beam propagation direction will move the cavity transmission within the spectral pit due to the small wedge on the crystal.
\par
After preparing the transparent (non-absorbing) spectral transmission window, each AFC peak is created using a complex hyperbolic secant pulse (sechyp for short) with the chirp width $f_{width}=70\;k\!H\!z$ \cite{Rippe2005} and temporal width $t_{FWHM}=16.8\,\mu s$ \cite{Rippe2005}. This pulse excites ions from $|\pm\frac{1}{2}g\rangle \mapsto |\pm\frac{5}{2}e\rangle$ state (See Fig. \ref{setup}a). From the $|\pm\frac{5}{2}e\rangle$ state, $Pr$ ions will decay mostly to the $|\pm\frac{5}{2}g\rangle$ due to the high branching ratio for the $|\pm\frac{5}{2}e\rangle \mapsto |\pm\frac{5}{2}g\rangle$ transition \cite{Nilsson2004}. This pulse is repeated several ($\sim 50$) times with a waiting time of $500\;\mu s$ in between each pulse. This process, creates one AFC peak. Repeating this procedure with a consecutive change of center frequency of the sechyp pulse by $\Delta$ will create the other AFC peaks.

\par
As we discussed earlier, the absorption engineering of the $Pr$ ions inside the cavity will directly affect the cavity modes via strong dispersion. Therefore, measuring the AFC structure properties in the cavity case is challenging. In order to at least to some extent estimate the AFC structure properties, a small part of the cavity crystal is left un-coated, and the same preparation as for the memory is performed in this part. The trace in Fig. \ref{Readout} is recorded using a weak readout pulse which is frequency chirped at a rate of $10\;kHz/\mu s$ across the frequency region of AFC structure. The overall spectral structure is complicated and a detailed discussion of the spectrum is beyond the scope of the present paper, however the storage is carried out using the four absorption peaks in the interval $-1.5\,M\!H\!z$ to $1.5\,M\!H\!z$ and the inset shows the spectral content of the storage pulse relative to these four peaks. The background absorption $d_0$ of the AFC structure in Fig. \ref{Readout} will affect the memory efficiency $\eta=\tilde{d}^2 exp(-\tilde{d})exp(\frac{-7}{F_{AFC}^2})exp(-d_0)$ where the effective absorption of the AFC structure is defined as $\tilde{d}=\frac{d-d_0}{F_{AFC}}$ where $d=\alpha L$ and $d_0$ is the background absorption \cite{Timoney2012}. To reduce $d_0$ and increase the overall quantum memory efficiency, sets of sechyp pulses with $f_{width}=300\;kHz$ and $t_{FWHM}=20\;\mu s$ \cite{Rippe2005} were used to scan and clean up the background absorption between the AFC peaks.

\begin{figure}[ht]
    \includegraphics[width=8cm]{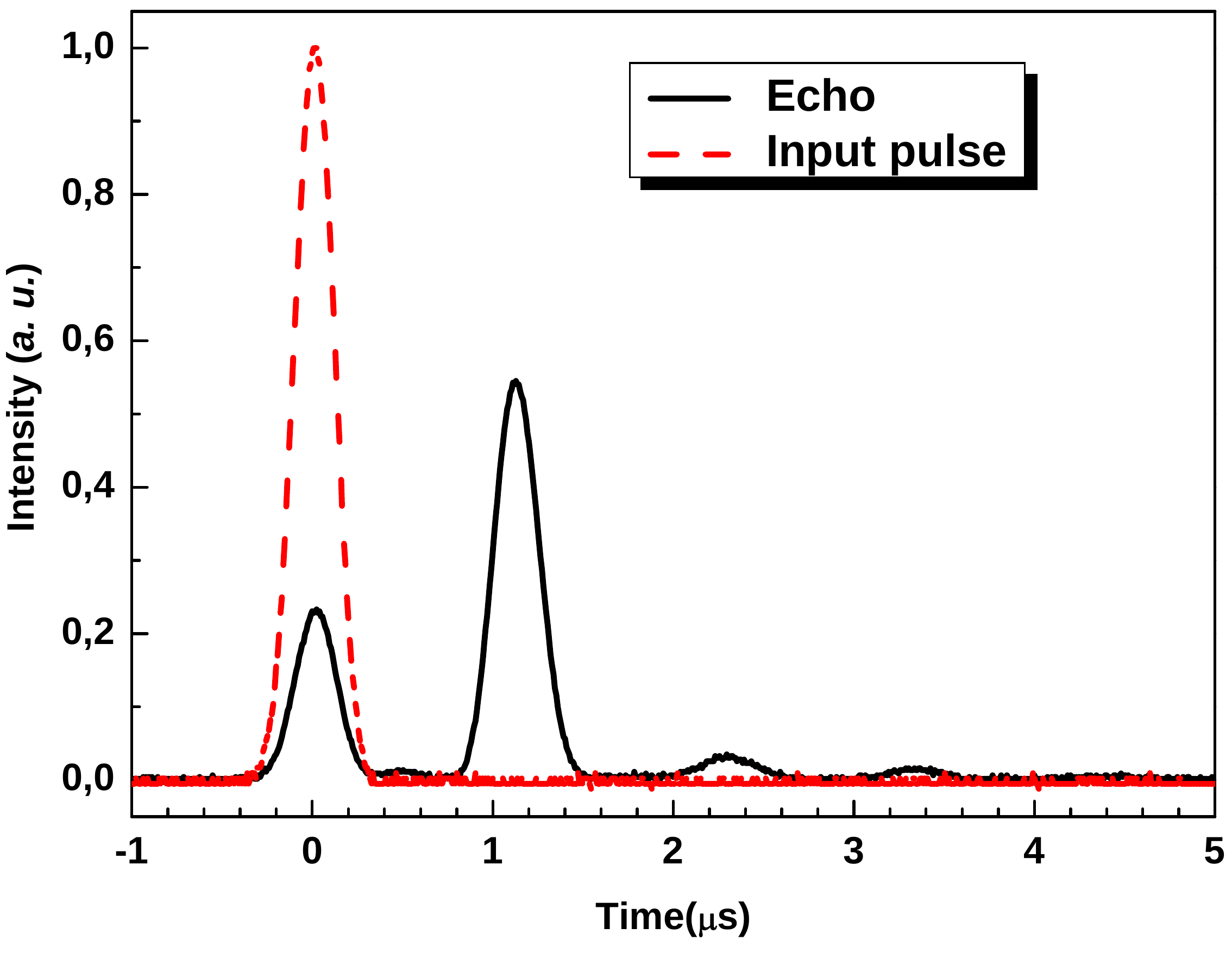}
    \caption{(Color online) The input storage pulse as a reference detected at PD3 (see text) is shown as a red dashed line. The retrieved echo pulse is detected at detector PD3 after $1.1\;\mu s$ as shown with the (black) solid line. The area of the echo pulse at $1.1\;\mu s$ divided by the reference signal pulse area gives a storage and retrieval efficiency of the memory of $\eta=58 \pm 5\%$.}
    \label{echo}
\end{figure}

\par
The input pulse is stored using the $|\pm\frac{5}{2}g\rangle \mapsto |\pm\frac{5}{2}e\rangle$ transition for ions initially in state $|\pm\frac{5}{2}g\rangle$. A Gaussian pulse with a duration of $\tau_{FWHM}=250\;ns$ and small pulse area is employed as a storage pulse. The frequency of the storage pulse is tuned to the center frequency of an AFC structure with a peak separation $\Delta=0.9\;MHz$. The retrieved echo pulse is detected at detector PD3 after $1.1\;\mu s$ as shown with black solid line in Fig. \ref{echo}.
\par
In order to assess the storage and retrieval efficiency, the intensity of the input storage pulse at the cavity crystal is used as a reference. To this end, the laser frequency was set to be off-resonance ($\approx 1\;nm$) with the inhomogeneous broadening of the $Pr$ ions. Several measurement above and below the inhomogeneous broadening center frequency is done. For comparison with off-line reference measurement the cavity crystal was turned $\sim\;180^{\circ}$ such that the input storage pulse impinged on the $R2=99.7\%$ mirror. In this way the input storage pulse is (almost completely) reflected and the signal on PD3 can be used as a reference value for the storage pulse input intensity. The reference measurements show a statistical variation of about $5\%$, which limits the precision of the memory efficiency measurement. The reference storage pulse detected at PD3 is shown as a red dashed line in Fig. \ref{echo}. The first echo pulse area divided by the reference signal pulse area gives a storage and retrieval efficiency of the memory of $\eta=58 \pm 5\%$.

\par
The present result is lower than the best storage and retrieval efficiency achieved elsewhere \cite{Hosseini2011,Hedges2010}, however it is the highest storage and retrieval efficiency based on the AFC protocol which is presently the best multi-mode quantum state storage protocol \cite{Nunn2008}. In order to obtain on-demand and long-term storage based on the AFC protocol \cite{Afzelius2009}, the ground-excited state superposition should be transferred to, and then brought back from, a spin-level superposition between two of the ground states \cite{Afzelius2010}. The de-phasing due to the spin state inhomogeneous broadening can be suppressed using radio-frequency (RF) spin-echo techniques. In addition, even longer ($> 60\,s$) storage time is possible by utilizing spin-echo techniques to suppress slow variations of the spin transition frequencies \cite{Longdell2005,Heinze2012}. Quantum memories demonstrated in a smaller physical volume would require significantly lower RF power. Although the present result is $10\%$ lower than the highest rare-earth crystal efficiency results so far \cite{Hedges2010}, it is obtained in a crystal which is 10 times shorter. This may in practice significantly simplify long time, high efficiency spin storage, since too large volumes will require excessive RF powers to compensate for the spin inhomogeneous broadening. In addition, efficient quantum memory in a weakly absorbing media, opens up the possibility of utilizing materials with low optical depth but good coherence properties (e.g. in $Eu^{3+}\!:\!Y_2SiO_5$).

\par
In summary, we have demonstrated a quantum memory with $\eta=58 \pm 5\%$ storage and retrieval efficiency based on the AFC protocol. This is done in a weakly absorbing medium and short crystal length ($2\,mm$) by utilizing an impedance-match cavity configuration. This achievement, in addition to the storage and recall of weak coherent optical pulses in Ref. \cite{Usmani2010,Sabooni2010}, spin-wave storage demonstration in Ref. \cite{Afzelius2010}, and the best multi-mode quantum memory in Ref. \cite{Bonarota2011,Usmani2010} proves the possibility of extending an efficient, on-demand, long storage time, and multi-mode quantum memory based on AFC protocol in the future.

\par
This work was supported by the Swedish Research Council, the Knut \& Alice Wallenberg Foundation, the Maja \& Erik Lindqvists forskningsstiftelse, the Crafoord Foundation and the EC FP7 Contract No. 247743 (QuRep) and (Marie Curie Action) REA grant agreement no. 287252 (CIPRIS).

\bibliography{C:/MAHMOOD/ms_thesis/References/MS_Ref}

\begin{thebibliography}{39}%
\makeatletter
\providecommand \@ifxundefined [1]{%
 \@ifx{#1\undefined}
}%
\providecommand \@ifnum [1]{%
 \ifnum #1\expandafter \@firstoftwo
 \else \expandafter \@secondoftwo
 \fi
}%
\providecommand \@ifx [1]{%
 \ifx #1\expandafter \@firstoftwo
 \else \expandafter \@secondoftwo
 \fi
}%
\providecommand \natexlab [1]{#1}%
\providecommand \enquote  [1]{``#1''}%
\providecommand \bibnamefont  [1]{#1}%
\providecommand \bibfnamefont [1]{#1}%
\providecommand \citenamefont [1]{#1}%
\providecommand \href@noop [0]{\@secondoftwo}%
\providecommand \href [0]{\begingroup \@sanitize@url \@href}%
\providecommand \@href[1]{\@@startlink{#1}\@@href}%
\providecommand \@@href[1]{\endgroup#1\@@endlink}%
\providecommand \@sanitize@url [0]{\catcode `\\12\catcode `\$12\catcode
  `\&12\catcode `\#12\catcode `\^12\catcode `\_12\catcode `\%12\relax}%
\providecommand \@@startlink[1]{}%
\providecommand \@@endlink[0]{}%
\providecommand \url  [0]{\begingroup\@sanitize@url \@url }%
\providecommand \@url [1]{\endgroup\@href {#1}{\urlprefix }}%
\providecommand \urlprefix  [0]{URL }%
\providecommand \Eprint [0]{\href }%
\providecommand \doibase [0]{http://dx.doi.org/}%
\providecommand \selectlanguage [0]{\@gobble}%
\providecommand \bibinfo  [0]{\@secondoftwo}%
\providecommand \bibfield  [0]{\@secondoftwo}%
\providecommand \translation [1]{[#1]}%
\providecommand \BibitemOpen [0]{}%
\providecommand \bibitemStop [0]{}%
\providecommand \bibitemNoStop [0]{.\EOS\space}%
\providecommand \EOS [0]{\spacefactor3000\relax}%
\providecommand \BibitemShut  [1]{\csname bibitem#1\endcsname}%
\let\auto@bib@innerbib\@empty
\bibitem [{\citenamefont {Hammerer}\ \emph {et~al.}(2010)\citenamefont
  {Hammerer}, \citenamefont {Sorensen},\ and\ \citenamefont
  {Polzik}}]{Hammerer2010}%
  \BibitemOpen
  \bibfield  {author} {\bibinfo {author} {\bibfnamefont {K.}~\bibnamefont
  {Hammerer}}, \bibinfo {author} {\bibfnamefont {A.~S.}\ \bibnamefont
  {Sorensen}}, \ and\ \bibinfo {author} {\bibfnamefont {E.~S.}\ \bibnamefont
  {Polzik}},\ }\href@noop {} {\bibfield  {journal} {\bibinfo  {journal}
  {Reviews of Modern Physics}\ }\textbf {\bibinfo {volume} {82}},\ \bibinfo
  {pages} {1041} (\bibinfo {year} {2010})}\BibitemShut {NoStop}%
\bibitem [{\citenamefont {Duan}\ \emph {et~al.}(2001)\citenamefont {Duan},
  \citenamefont {Lukin}, \citenamefont {Cirac},\ and\ \citenamefont
  {Zoller}}]{Duan2001}%
  \BibitemOpen
  \bibfield  {author} {\bibinfo {author} {\bibfnamefont {L.~M.}\ \bibnamefont
  {Duan}}, \bibinfo {author} {\bibfnamefont {M.~D.}\ \bibnamefont {Lukin}},
  \bibinfo {author} {\bibfnamefont {J.~I.}\ \bibnamefont {Cirac}}, \ and\
  \bibinfo {author} {\bibfnamefont {P.}~\bibnamefont {Zoller}},\ }\href@noop {}
  {\bibfield  {journal} {\bibinfo  {journal} {Nature}\ }\textbf {\bibinfo
  {volume} {414}},\ \bibinfo {pages} {413} (\bibinfo {year}
  {2001})}\BibitemShut {NoStop}%
\bibitem [{\citenamefont {Sangouard}\ \emph {et~al.}(2008)\citenamefont
  {Sangouard}, \citenamefont {Simon}, \citenamefont {Zhao}, \citenamefont
  {Chen}, \citenamefont {de~Riedmatten}, \citenamefont {Pan},\ and\
  \citenamefont {Gisin}}]{Sangouard2008}%
  \BibitemOpen
  \bibfield  {author} {\bibinfo {author} {\bibfnamefont {N.}~\bibnamefont
  {Sangouard}}, \bibinfo {author} {\bibfnamefont {C.}~\bibnamefont {Simon}},
  \bibinfo {author} {\bibfnamefont {B.}~\bibnamefont {Zhao}}, \bibinfo {author}
  {\bibfnamefont {Y.~A.}\ \bibnamefont {Chen}}, \bibinfo {author}
  {\bibfnamefont {H.}~\bibnamefont {de~Riedmatten}}, \bibinfo {author}
  {\bibfnamefont {J.~W.}\ \bibnamefont {Pan}}, \ and\ \bibinfo {author}
  {\bibfnamefont {N.}~\bibnamefont {Gisin}},\ }\href@noop {} {\bibfield
  {journal} {\bibinfo  {journal} {Physical Review A}\ }\textbf {\bibinfo
  {volume} {77}},\ \bibinfo {pages} {062301} (\bibinfo {year}
  {2008})}\BibitemShut {NoStop}%
\bibitem [{\citenamefont {Kok}\ \emph {et~al.}(2007)\citenamefont {Kok},
  \citenamefont {Munro}, \citenamefont {Nemoto}, \citenamefont {Ralph},
  \citenamefont {Dowling},\ and\ \citenamefont {Milburn}}]{Kok2007}%
  \BibitemOpen
  \bibfield  {author} {\bibinfo {author} {\bibfnamefont {P.}~\bibnamefont
  {Kok}}, \bibinfo {author} {\bibfnamefont {W.~J.}\ \bibnamefont {Munro}},
  \bibinfo {author} {\bibfnamefont {K.}~\bibnamefont {Nemoto}}, \bibinfo
  {author} {\bibfnamefont {T.~C.}\ \bibnamefont {Ralph}}, \bibinfo {author}
  {\bibfnamefont {J.~P.}\ \bibnamefont {Dowling}}, \ and\ \bibinfo {author}
  {\bibfnamefont {G.~J.}\ \bibnamefont {Milburn}},\ }\href@noop {} {\bibfield
  {journal} {\bibinfo  {journal} {Reviews of Modern Physics}\ }\textbf
  {\bibinfo {volume} {79}},\ \bibinfo {pages} {135} (\bibinfo {year}
  {2007})}\BibitemShut {NoStop}%
\bibitem [{\citenamefont {Knill}\ \emph {et~al.}(2001)\citenamefont {Knill},
  \citenamefont {Laflamme},\ and\ \citenamefont {Milburn}}]{Knill2001}%
  \BibitemOpen
  \bibfield  {author} {\bibinfo {author} {\bibfnamefont {E.}~\bibnamefont
  {Knill}}, \bibinfo {author} {\bibfnamefont {R.}~\bibnamefont {Laflamme}}, \
  and\ \bibinfo {author} {\bibfnamefont {G.~J.}\ \bibnamefont {Milburn}},\
  }\href@noop {} {\bibfield  {journal} {\bibinfo  {journal} {Nature}\ }\textbf
  {\bibinfo {volume} {409}},\ \bibinfo {pages} {46} (\bibinfo {year}
  {2001})}\BibitemShut {NoStop}%
\bibitem [{\citenamefont {Nunn}\ \emph {et~al.}(2012)\citenamefont {Nunn},
  \citenamefont {Langford}, \citenamefont {Kolthammer}, \citenamefont
  {Champion}, \citenamefont {Sprague}, \citenamefont {Michelberger},
  \citenamefont {Jin}, \citenamefont {England},\ and\ \citenamefont
  {Walmsley}}]{Nunn2012}%
  \BibitemOpen
  \bibfield  {author} {\bibinfo {author} {\bibfnamefont {J.}~\bibnamefont
  {Nunn}}, \bibinfo {author} {\bibfnamefont {N.~K.}\ \bibnamefont {Langford}},
  \bibinfo {author} {\bibfnamefont {W.~S.}\ \bibnamefont {Kolthammer}},
  \bibinfo {author} {\bibfnamefont {T.~F.~M.}\ \bibnamefont {Champion}},
  \bibinfo {author} {\bibfnamefont {M.~R.}\ \bibnamefont {Sprague}}, \bibinfo
  {author} {\bibfnamefont {P.~S.}\ \bibnamefont {Michelberger}}, \bibinfo
  {author} {\bibfnamefont {X.~M.}\ \bibnamefont {Jin}}, \bibinfo {author}
  {\bibfnamefont {D.~G.}\ \bibnamefont {England}}, \ and\ \bibinfo {author}
  {\bibfnamefont {I.~A.}\ \bibnamefont {Walmsley}},\ }\href@noop {} {\bibfield
  {journal} {\bibinfo  {journal} {arXiv:1208.1534v1}\ } (\bibinfo {year}
  {2012})}\BibitemShut {NoStop}%
\bibitem [{\citenamefont {Chen}\ \emph {et~al.}(2006)\citenamefont {Chen},
  \citenamefont {Chen}, \citenamefont {Strassel}, \citenamefont {Yuan},
  \citenamefont {Zhao}, \citenamefont {Schmiedmayer},\ and\ \citenamefont
  {Pan}}]{Chen2006}%
  \BibitemOpen
  \bibfield  {author} {\bibinfo {author} {\bibfnamefont {S.}~\bibnamefont
  {Chen}}, \bibinfo {author} {\bibfnamefont {Y.-A.}\ \bibnamefont {Chen}},
  \bibinfo {author} {\bibfnamefont {T.}~\bibnamefont {Strassel}}, \bibinfo
  {author} {\bibfnamefont {Z.-S.}\ \bibnamefont {Yuan}}, \bibinfo {author}
  {\bibfnamefont {B.}~\bibnamefont {Zhao}}, \bibinfo {author} {\bibfnamefont
  {J.}~\bibnamefont {Schmiedmayer}}, \ and\ \bibinfo {author} {\bibfnamefont
  {J.-W.}\ \bibnamefont {Pan}},\ }\href@noop {} {\bibfield  {journal} {\bibinfo
   {journal} {Physical Review Letters}\ }\textbf {\bibinfo {volume} {97}},\
  \bibinfo {pages} {173004} (\bibinfo {year} {2006})}\BibitemShut {NoStop}%
\bibitem [{\citenamefont {Lvovsky}\ \emph {et~al.}(2009)\citenamefont
  {Lvovsky}, \citenamefont {Sanders},\ and\ \citenamefont
  {Tittel}}]{Lvovsky2009}%
  \BibitemOpen
  \bibfield  {author} {\bibinfo {author} {\bibfnamefont {A.~I.}\ \bibnamefont
  {Lvovsky}}, \bibinfo {author} {\bibfnamefont {B.~C.}\ \bibnamefont
  {Sanders}}, \ and\ \bibinfo {author} {\bibfnamefont {W.}~\bibnamefont
  {Tittel}},\ }\href@noop {} {\bibfield  {journal} {\bibinfo  {journal} {Nature
  Photonics}\ }\textbf {\bibinfo {volume} {3}},\ \bibinfo {pages} {706}
  (\bibinfo {year} {2009})}\BibitemShut {NoStop}%
\bibitem [{\citenamefont {Hosseini}\ \emph {et~al.}(2011)\citenamefont
  {Hosseini}, \citenamefont {Sparkes}, \citenamefont {Campbell}, \citenamefont
  {Lam},\ and\ \citenamefont {Buchler}}]{Hosseini2011}%
  \BibitemOpen
  \bibfield  {author} {\bibinfo {author} {\bibfnamefont {M.}~\bibnamefont
  {Hosseini}}, \bibinfo {author} {\bibfnamefont {B.~M.}\ \bibnamefont
  {Sparkes}}, \bibinfo {author} {\bibfnamefont {G.}~\bibnamefont {Campbell}},
  \bibinfo {author} {\bibfnamefont {P.~K.}\ \bibnamefont {Lam}}, \ and\
  \bibinfo {author} {\bibfnamefont {B.~C.}\ \bibnamefont {Buchler}},\
  }\href@noop {} {\bibfield  {journal} {\bibinfo  {journal} {Nature
  Communications}\ }\textbf {\bibinfo {volume} {2}},\ \bibinfo {pages} {174}
  (\bibinfo {year} {2011})}\BibitemShut {NoStop}%
\bibitem [{\citenamefont {Hedges}\ \emph {et~al.}(2010)\citenamefont {Hedges},
  \citenamefont {Longdell}, \citenamefont {Li},\ and\ \citenamefont
  {Sellars}}]{Hedges2010}%
  \BibitemOpen
  \bibfield  {author} {\bibinfo {author} {\bibfnamefont {M.~P.}\ \bibnamefont
  {Hedges}}, \bibinfo {author} {\bibfnamefont {J.~J.}\ \bibnamefont
  {Longdell}}, \bibinfo {author} {\bibfnamefont {Y.}~\bibnamefont {Li}}, \ and\
  \bibinfo {author} {\bibfnamefont {M.~J.}\ \bibnamefont {Sellars}},\
  }\href@noop {} {\bibfield  {journal} {\bibinfo  {journal} {Nature}\ }\textbf
  {\bibinfo {volume} {465}},\ \bibinfo {pages} {1052} (\bibinfo {year}
  {2010})}\BibitemShut {NoStop}%
\bibitem [{\citenamefont {Longdell}\ \emph {et~al.}(2005)\citenamefont
  {Longdell}, \citenamefont {Fraval}, \citenamefont {Sellars},\ and\
  \citenamefont {Manson}}]{Longdell2005}%
  \BibitemOpen
  \bibfield  {author} {\bibinfo {author} {\bibfnamefont {J.~J.}\ \bibnamefont
  {Longdell}}, \bibinfo {author} {\bibfnamefont {E.}~\bibnamefont {Fraval}},
  \bibinfo {author} {\bibfnamefont {M.~J.}\ \bibnamefont {Sellars}}, \ and\
  \bibinfo {author} {\bibfnamefont {N.~B.}\ \bibnamefont {Manson}},\
  }\href@noop {} {\bibfield  {journal} {\bibinfo  {journal} {Physical Review
  Letters}\ }\textbf {\bibinfo {volume} {95}},\ \bibinfo {pages} {063601}
  (\bibinfo {year} {2005})}\BibitemShut {NoStop}%
\bibitem [{\citenamefont {Heinze}(2012)}]{Heinze2012}%
  \BibitemOpen
  \bibfield  {author} {\bibinfo {author} {\bibfnamefont {G.}~\bibnamefont
  {Heinze}},\ }\href@noop {} {\bibfield  {journal} {\bibinfo  {journal} {et.
  al. (to be published).}\ } (\bibinfo {year} {2012})}\BibitemShut {NoStop}%
\bibitem [{\citenamefont {Bonarota}\ \emph {et~al.}(2011)\citenamefont
  {Bonarota}, \citenamefont {Le~Gou$\ddot{e}$t},\ and\ \citenamefont
  {Chaneli$\grave{e}$re}}]{Bonarota2011}%
  \BibitemOpen
  \bibfield  {author} {\bibinfo {author} {\bibfnamefont {M.}~\bibnamefont
  {Bonarota}}, \bibinfo {author} {\bibfnamefont {J.~L.}\ \bibnamefont
  {Le~Gou$\ddot{e}$t}}, \ and\ \bibinfo {author} {\bibfnamefont
  {T.}~\bibnamefont {Chaneli$\grave{e}$re}},\ }\href@noop {} {\bibfield
  {journal} {\bibinfo  {journal} {New Journal of Physics}\ }\textbf {\bibinfo
  {volume} {13}},\ \bibinfo {pages} {013013} (\bibinfo {year}
  {2011})}\BibitemShut {NoStop}%
\bibitem [{\citenamefont {Usmani}\ \emph {et~al.}(2010)\citenamefont {Usmani},
  \citenamefont {Afzelius}, \citenamefont {de~Riedmatten},\ and\ \citenamefont
  {Gisin}}]{Usmani2010}%
  \BibitemOpen
  \bibfield  {author} {\bibinfo {author} {\bibfnamefont {I.}~\bibnamefont
  {Usmani}}, \bibinfo {author} {\bibfnamefont {M.}~\bibnamefont {Afzelius}},
  \bibinfo {author} {\bibfnamefont {H.}~\bibnamefont {de~Riedmatten}}, \ and\
  \bibinfo {author} {\bibfnamefont {N.}~\bibnamefont {Gisin}},\ }\href@noop {}
  {\bibfield  {journal} {\bibinfo  {journal} {Nature Communications}\ }\textbf
  {\bibinfo {volume} {1}},\ \bibinfo {pages} {12} (\bibinfo {year}
  {2010})}\BibitemShut {NoStop}%
\bibitem [{\citenamefont {Saglamyurek}\ \emph {et~al.}(2011)\citenamefont
  {Saglamyurek}, \citenamefont {Sinclair}, \citenamefont {Jin}, \citenamefont
  {Slater}, \citenamefont {Oblak}, \citenamefont {Bussieres}, \citenamefont
  {George}, \citenamefont {Ricken}, \citenamefont {Sohler},\ and\ \citenamefont
  {Tittel}}]{Saglamyurek2011}%
  \BibitemOpen
  \bibfield  {author} {\bibinfo {author} {\bibfnamefont {E.}~\bibnamefont
  {Saglamyurek}}, \bibinfo {author} {\bibfnamefont {N.}~\bibnamefont
  {Sinclair}}, \bibinfo {author} {\bibfnamefont {J.}~\bibnamefont {Jin}},
  \bibinfo {author} {\bibfnamefont {J.~A.}\ \bibnamefont {Slater}}, \bibinfo
  {author} {\bibfnamefont {D.}~\bibnamefont {Oblak}}, \bibinfo {author}
  {\bibfnamefont {F.}~\bibnamefont {Bussieres}}, \bibinfo {author}
  {\bibfnamefont {M.}~\bibnamefont {George}}, \bibinfo {author} {\bibfnamefont
  {R.}~\bibnamefont {Ricken}}, \bibinfo {author} {\bibfnamefont
  {W.}~\bibnamefont {Sohler}}, \ and\ \bibinfo {author} {\bibfnamefont
  {W.}~\bibnamefont {Tittel}},\ }\href@noop {} {\bibfield  {journal} {\bibinfo
  {journal} {Nature}\ }\textbf {\bibinfo {volume} {469}},\ \bibinfo {pages}
  {512} (\bibinfo {year} {2011})}\BibitemShut {NoStop}%
\bibitem [{\citenamefont {Fleischhauer}\ and\ \citenamefont
  {Lukin}(2000)}]{Fleischhauer2000a}%
  \BibitemOpen
  \bibfield  {author} {\bibinfo {author} {\bibfnamefont {M.}~\bibnamefont
  {Fleischhauer}}\ and\ \bibinfo {author} {\bibfnamefont {M.~D.}\ \bibnamefont
  {Lukin}},\ }\href@noop {} {\bibfield  {journal} {\bibinfo  {journal}
  {Physical Review Letters}\ }\textbf {\bibinfo {volume} {84}},\ \bibinfo
  {pages} {5094} (\bibinfo {year} {2000})}\BibitemShut {NoStop}%
\bibitem [{\citenamefont {Reim}\ \emph {et~al.}(2010)\citenamefont {Reim},
  \citenamefont {Nunn}, \citenamefont {Lorenz}, \citenamefont {Sussman},
  \citenamefont {Lee}, \citenamefont {Langford}, \citenamefont {Jaksch},\ and\
  \citenamefont {Walmsley}}]{Reim2010}%
  \BibitemOpen
  \bibfield  {author} {\bibinfo {author} {\bibfnamefont {K.~F.}\ \bibnamefont
  {Reim}}, \bibinfo {author} {\bibfnamefont {J.}~\bibnamefont {Nunn}}, \bibinfo
  {author} {\bibfnamefont {V.~O.}\ \bibnamefont {Lorenz}}, \bibinfo {author}
  {\bibfnamefont {B.~J.}\ \bibnamefont {Sussman}}, \bibinfo {author}
  {\bibfnamefont {K.~C.}\ \bibnamefont {Lee}}, \bibinfo {author} {\bibfnamefont
  {N.~K.}\ \bibnamefont {Langford}}, \bibinfo {author} {\bibfnamefont
  {D.}~\bibnamefont {Jaksch}}, \ and\ \bibinfo {author} {\bibfnamefont {I.~A.}\
  \bibnamefont {Walmsley}},\ }\href@noop {} {\bibfield  {journal} {\bibinfo
  {journal} {Nature Photonics}\ }\textbf {\bibinfo {volume} {4}},\ \bibinfo
  {pages} {218} (\bibinfo {year} {2010})}\BibitemShut {NoStop}%
\bibitem [{\citenamefont {Moiseev}\ and\ \citenamefont
  {Kr\"{o}ll}(2001)}]{Moiseev2001}%
  \BibitemOpen
  \bibfield  {author} {\bibinfo {author} {\bibfnamefont {S.~A.}\ \bibnamefont
  {Moiseev}}\ and\ \bibinfo {author} {\bibfnamefont {S.}~\bibnamefont
  {Kr\"{o}ll}},\ }\href@noop {} {\bibfield  {journal} {\bibinfo  {journal}
  {Physical Review Letters}\ }\textbf {\bibinfo {volume} {87}},\ \bibinfo
  {pages} {173601} (\bibinfo {year} {2001})}\BibitemShut {NoStop}%
\bibitem [{\citenamefont {Nilsson}\ and\ \citenamefont
  {Kr\"{o}ll}(2005)}]{Nilsson2005}%
  \BibitemOpen
  \bibfield  {author} {\bibinfo {author} {\bibfnamefont {M.}~\bibnamefont
  {Nilsson}}\ and\ \bibinfo {author} {\bibfnamefont {S.}~\bibnamefont
  {Kr\"{o}ll}},\ }\href@noop {} {\bibfield  {journal} {\bibinfo  {journal}
  {Optics Communications}\ }\textbf {\bibinfo {volume} {247}},\ \bibinfo
  {pages} {393} (\bibinfo {year} {2005})}\BibitemShut {NoStop}%
\bibitem [{\citenamefont {Kraus}\ \emph {et~al.}(2006)\citenamefont {Kraus},
  \citenamefont {Tittel}, \citenamefont {Gisin}, \citenamefont {Nilsson},
  \citenamefont {Kr\"{o}ll},\ and\ \citenamefont {Cirac}}]{Kraus2006}%
  \BibitemOpen
  \bibfield  {author} {\bibinfo {author} {\bibfnamefont {B.}~\bibnamefont
  {Kraus}}, \bibinfo {author} {\bibfnamefont {W.}~\bibnamefont {Tittel}},
  \bibinfo {author} {\bibfnamefont {N.}~\bibnamefont {Gisin}}, \bibinfo
  {author} {\bibfnamefont {M.}~\bibnamefont {Nilsson}}, \bibinfo {author}
  {\bibfnamefont {S.}~\bibnamefont {Kr\"{o}ll}}, \ and\ \bibinfo {author}
  {\bibfnamefont {J.~I.}\ \bibnamefont {Cirac}},\ }\href@noop {} {\bibfield
  {journal} {\bibinfo  {journal} {Physical Review A}\ }\textbf {\bibinfo
  {volume} {73}},\ \bibinfo {pages} {020302} (\bibinfo {year}
  {2006})}\BibitemShut {NoStop}%
\bibitem [{\citenamefont {Hetet}\ \emph {et~al.}(2008)\citenamefont {Hetet},
  \citenamefont {Longdell}, \citenamefont {Sellars}, \citenamefont {Lam},\ and\
  \citenamefont {Buchler}}]{Hetet2008a}%
  \BibitemOpen
  \bibfield  {author} {\bibinfo {author} {\bibfnamefont {G.}~\bibnamefont
  {Hetet}}, \bibinfo {author} {\bibfnamefont {J.~J.}\ \bibnamefont {Longdell}},
  \bibinfo {author} {\bibfnamefont {M.~J.}\ \bibnamefont {Sellars}}, \bibinfo
  {author} {\bibfnamefont {P.~K.}\ \bibnamefont {Lam}}, \ and\ \bibinfo
  {author} {\bibfnamefont {B.~C.}\ \bibnamefont {Buchler}},\ }\href@noop {}
  {\bibfield  {journal} {\bibinfo  {journal} {Physical Review Letters}\
  }\textbf {\bibinfo {volume} {101}},\ \bibinfo {pages} {203601} (\bibinfo
  {year} {2008})}\BibitemShut {NoStop}%
\bibitem [{\citenamefont {Afzelius}\ \emph {et~al.}(2009)\citenamefont
  {Afzelius}, \citenamefont {Simon}, \citenamefont {de~Riedmatten},\ and\
  \citenamefont {Gisin}}]{Afzelius2009}%
  \BibitemOpen
  \bibfield  {author} {\bibinfo {author} {\bibfnamefont {M.}~\bibnamefont
  {Afzelius}}, \bibinfo {author} {\bibfnamefont {C.}~\bibnamefont {Simon}},
  \bibinfo {author} {\bibfnamefont {H.}~\bibnamefont {de~Riedmatten}}, \ and\
  \bibinfo {author} {\bibfnamefont {N.}~\bibnamefont {Gisin}},\ }\href@noop {}
  {\bibfield  {journal} {\bibinfo  {journal} {Physical Review A}\ }\textbf
  {\bibinfo {volume} {79}},\ \bibinfo {pages} {052329} (\bibinfo {year}
  {2009})}\BibitemShut {NoStop}%
\bibitem [{\citenamefont {Phillips}\ \emph {et~al.}(2008)\citenamefont
  {Phillips}, \citenamefont {Gorshkov},\ and\ \citenamefont
  {Novikova}}]{Phillips2008}%
  \BibitemOpen
  \bibfield  {author} {\bibinfo {author} {\bibfnamefont {N.~B.}\ \bibnamefont
  {Phillips}}, \bibinfo {author} {\bibfnamefont {A.~V.}\ \bibnamefont
  {Gorshkov}}, \ and\ \bibinfo {author} {\bibfnamefont {I.}~\bibnamefont
  {Novikova}},\ }\href@noop {} {\bibfield  {journal} {\bibinfo  {journal}
  {Physical Review A}\ }\textbf {\bibinfo {volume} {78}},\ \bibinfo {pages}
  {023801} (\bibinfo {year} {2008})}\BibitemShut {NoStop}%
\bibitem [{\citenamefont {Amari}\ \emph {et~al.}(2010)\citenamefont {Amari},
  \citenamefont {Walther}, \citenamefont {Sabooni}, \citenamefont {Huang},
  \citenamefont {Kr\"{o}ll}, \citenamefont {Afzelius}, \citenamefont {Usmani},
  \citenamefont {Lauritzen}, \citenamefont {Sangouard}, \citenamefont
  {de~Riedmatten},\ and\ \citenamefont {Gisin}}]{Amari2010}%
  \BibitemOpen
  \bibfield  {author} {\bibinfo {author} {\bibfnamefont {A.}~\bibnamefont
  {Amari}}, \bibinfo {author} {\bibfnamefont {A.}~\bibnamefont {Walther}},
  \bibinfo {author} {\bibfnamefont {M.}~\bibnamefont {Sabooni}}, \bibinfo
  {author} {\bibfnamefont {M.}~\bibnamefont {Huang}}, \bibinfo {author}
  {\bibfnamefont {S.}~\bibnamefont {Kr\"{o}ll}}, \bibinfo {author}
  {\bibfnamefont {M.}~\bibnamefont {Afzelius}}, \bibinfo {author}
  {\bibfnamefont {I.}~\bibnamefont {Usmani}}, \bibinfo {author} {\bibfnamefont
  {B.}~\bibnamefont {Lauritzen}}, \bibinfo {author} {\bibfnamefont
  {N.}~\bibnamefont {Sangouard}}, \bibinfo {author} {\bibfnamefont
  {H.}~\bibnamefont {de~Riedmatten}}, \ and\ \bibinfo {author} {\bibfnamefont
  {N.}~\bibnamefont {Gisin}},\ }\href@noop {} {\bibfield  {journal} {\bibinfo
  {journal} {Journal of Luminescence}\ }\textbf {\bibinfo {volume} {130}},\
  \bibinfo {pages} {1579} (\bibinfo {year} {2010})}\BibitemShut {NoStop}%
\bibitem [{\citenamefont {Afzelius}\ \emph {et~al.}(2010)\citenamefont
  {Afzelius}, \citenamefont {Usmani}, \citenamefont {Amari}, \citenamefont
  {Lauritzen}, \citenamefont {Walther}, \citenamefont {Simon}, \citenamefont
  {Sangouard}, \citenamefont {Minar}, \citenamefont {de~Riedmatten},
  \citenamefont {Gisin},\ and\ \citenamefont {Kr\"{o}ll}}]{Afzelius2010}%
  \BibitemOpen
  \bibfield  {author} {\bibinfo {author} {\bibfnamefont {M.}~\bibnamefont
  {Afzelius}}, \bibinfo {author} {\bibfnamefont {I.}~\bibnamefont {Usmani}},
  \bibinfo {author} {\bibfnamefont {A.}~\bibnamefont {Amari}}, \bibinfo
  {author} {\bibfnamefont {B.}~\bibnamefont {Lauritzen}}, \bibinfo {author}
  {\bibfnamefont {A.}~\bibnamefont {Walther}}, \bibinfo {author} {\bibfnamefont
  {C.}~\bibnamefont {Simon}}, \bibinfo {author} {\bibfnamefont
  {N.}~\bibnamefont {Sangouard}}, \bibinfo {author} {\bibfnamefont
  {J.}~\bibnamefont {Minar}}, \bibinfo {author} {\bibfnamefont
  {H.}~\bibnamefont {de~Riedmatten}}, \bibinfo {author} {\bibfnamefont
  {N.}~\bibnamefont {Gisin}}, \ and\ \bibinfo {author} {\bibfnamefont
  {S.}~\bibnamefont {Kr\"{o}ll}},\ }\href@noop {} {\bibfield  {journal}
  {\bibinfo  {journal} {Physical Review Letters}\ }\textbf {\bibinfo {volume}
  {104}},\ \bibinfo {pages} {040503} (\bibinfo {year} {2010})}\BibitemShut
  {NoStop}%
\bibitem [{\citenamefont {Raimond}\ \emph {et~al.}(2001)\citenamefont
  {Raimond}, \citenamefont {Brune},\ and\ \citenamefont
  {Haroche}}]{Raimond2001}%
  \BibitemOpen
  \bibfield  {author} {\bibinfo {author} {\bibfnamefont {J.~M.}\ \bibnamefont
  {Raimond}}, \bibinfo {author} {\bibfnamefont {M.}~\bibnamefont {Brune}}, \
  and\ \bibinfo {author} {\bibfnamefont {S.}~\bibnamefont {Haroche}},\
  }\href@noop {} {\bibfield  {journal} {\bibinfo  {journal} {Reviews of Modern
  Physics}\ }\textbf {\bibinfo {volume} {73}},\ \bibinfo {pages} {565}
  (\bibinfo {year} {2001})}\BibitemShut {NoStop}%
\bibitem [{\citenamefont {Julsgaard}\ \emph {et~al.}(2004)\citenamefont
  {Julsgaard}, \citenamefont {Sherson}, \citenamefont {Cirac}, \citenamefont
  {Fiurasek},\ and\ \citenamefont {Polzik}}]{Julsgaard2004}%
  \BibitemOpen
  \bibfield  {author} {\bibinfo {author} {\bibfnamefont {B.}~\bibnamefont
  {Julsgaard}}, \bibinfo {author} {\bibfnamefont {J.}~\bibnamefont {Sherson}},
  \bibinfo {author} {\bibfnamefont {J.~I.}\ \bibnamefont {Cirac}}, \bibinfo
  {author} {\bibfnamefont {J.}~\bibnamefont {Fiurasek}}, \ and\ \bibinfo
  {author} {\bibfnamefont {E.~S.}\ \bibnamefont {Polzik}},\ }\href@noop {}
  {\bibfield  {journal} {\bibinfo  {journal} {Nature}\ }\textbf {\bibinfo
  {volume} {432}},\ \bibinfo {pages} {482} (\bibinfo {year}
  {2004})}\BibitemShut {NoStop}%
\bibitem [{\citenamefont {Simon}\ \emph {et~al.}(2007)\citenamefont {Simon},
  \citenamefont {Tanji}, \citenamefont {Thompson},\ and\ \citenamefont
  {Vuletic}}]{Simon2007a}%
  \BibitemOpen
  \bibfield  {author} {\bibinfo {author} {\bibfnamefont {J.}~\bibnamefont
  {Simon}}, \bibinfo {author} {\bibfnamefont {H.}~\bibnamefont {Tanji}},
  \bibinfo {author} {\bibfnamefont {J.~K.}\ \bibnamefont {Thompson}}, \ and\
  \bibinfo {author} {\bibfnamefont {V.}~\bibnamefont {Vuletic}},\ }\href@noop
  {} {\bibfield  {journal} {\bibinfo  {journal} {Physical Review Letters}\
  }\textbf {\bibinfo {volume} {98}},\ \bibinfo {pages} {183601} (\bibinfo
  {year} {2007})}\BibitemShut {NoStop}%
\bibitem [{\citenamefont {Liu}\ and\ \citenamefont
  {Jacquier}(2005)}]{GuokuiLiu2005}%
  \BibitemOpen
  \bibfield  {author} {\bibinfo {author} {\bibfnamefont {G.}~\bibnamefont
  {Liu}}\ and\ \bibinfo {author} {\bibfnamefont {B.}~\bibnamefont {Jacquier}},\
  }\href@noop {} {\emph {\bibinfo {title} {Spectroscopic Properties of Rare
  Earths in Optical Materials, Chapter 7}}}\ (\bibinfo  {publisher} {Springer
  Series in Material Science},\ \bibinfo {year} {2005})\BibitemShut {NoStop}%
\bibitem [{\citenamefont {Rippe}\ \emph {et~al.}(2005)\citenamefont {Rippe},
  \citenamefont {Nilsson}, \citenamefont {Kr\"{o}ll}, \citenamefont {Klieber},\
  and\ \citenamefont {Suter}}]{Rippe2005}%
  \BibitemOpen
  \bibfield  {author} {\bibinfo {author} {\bibfnamefont {L.}~\bibnamefont
  {Rippe}}, \bibinfo {author} {\bibfnamefont {M.}~\bibnamefont {Nilsson}},
  \bibinfo {author} {\bibfnamefont {S.}~\bibnamefont {Kr\"{o}ll}}, \bibinfo
  {author} {\bibfnamefont {R.}~\bibnamefont {Klieber}}, \ and\ \bibinfo
  {author} {\bibfnamefont {D.}~\bibnamefont {Suter}},\ }\href@noop {}
  {\bibfield  {journal} {\bibinfo  {journal} {Physical Review A}\ }\textbf
  {\bibinfo {volume} {71}},\ \bibinfo {pages} {062328} (\bibinfo {year}
  {2005})}\BibitemShut {NoStop}%
\bibitem [{\citenamefont {Afzelius}\ and\ \citenamefont
  {Simon}(2010)}]{Afzelius2010a}%
  \BibitemOpen
  \bibfield  {author} {\bibinfo {author} {\bibfnamefont {M.}~\bibnamefont
  {Afzelius}}\ and\ \bibinfo {author} {\bibfnamefont {C.}~\bibnamefont
  {Simon}},\ }\href@noop {} {\bibfield  {journal} {\bibinfo  {journal}
  {Physical Review A}\ }\textbf {\bibinfo {volume} {82}},\ \bibinfo {pages}
  {022310} (\bibinfo {year} {2010})}\BibitemShut {NoStop}%
\bibitem [{\citenamefont {Moiseev}\ \emph {et~al.}(2010)\citenamefont
  {Moiseev}, \citenamefont {Andrianov},\ and\ \citenamefont
  {Gubaidullin}}]{Moiseev2010}%
  \BibitemOpen
  \bibfield  {author} {\bibinfo {author} {\bibfnamefont {S.~A.}\ \bibnamefont
  {Moiseev}}, \bibinfo {author} {\bibfnamefont {S.~N.}\ \bibnamefont
  {Andrianov}}, \ and\ \bibinfo {author} {\bibfnamefont {F.~F.}\ \bibnamefont
  {Gubaidullin}},\ }\href@noop {} {\bibfield  {journal} {\bibinfo  {journal}
  {Physical Review A}\ }\textbf {\bibinfo {volume} {82}},\ \bibinfo {pages}
  {022311} (\bibinfo {year} {2010})}\BibitemShut {NoStop}%
\bibitem [{\citenamefont {Walther}\ \emph {et~al.}(ects)\citenamefont
  {Walther}, \citenamefont {Amari}, \citenamefont {Kr\"{o}ll},\ and\
  \citenamefont {Kalachev}}]{Walther2009a}%
  \BibitemOpen
  \bibfield  {author} {\bibinfo {author} {\bibfnamefont {A.}~\bibnamefont
  {Walther}}, \bibinfo {author} {\bibfnamefont {A.}~\bibnamefont {Amari}},
  \bibinfo {author} {\bibfnamefont {S.}~\bibnamefont {Kr\"{o}ll}}, \ and\
  \bibinfo {author} {\bibfnamefont {A.}~\bibnamefont {Kalachev}},\ }\href@noop
  {} {\bibfield  {journal} {\bibinfo  {journal} {Physical Review A}\ }\textbf
  {\bibinfo {volume} {80}},\ \bibinfo {pages} {012317} (\bibinfo {year}
  {2009).( Correction: the expression for the group velocity needs to be
  corrected to $v_g=\frac{2\pi\Gamma}{\alpha}$ in subsection "IV.B. Slow light
  effects")})}\BibitemShut {NoStop}%
\bibitem [{\citenamefont {Siegman}(1985)}]{Siegman1985}%
  \BibitemOpen
  \bibfield  {author} {\bibinfo {author} {\bibfnamefont {A.~E.}\ \bibnamefont
  {Siegman}},\ }\href@noop {} {\emph {\bibinfo {title} {Lasers}}}\ (\bibinfo
  {publisher} {University Science Books, Mill Valley, California},\ \bibinfo
  {year} {1985})\BibitemShut {NoStop}%
\bibitem [{\citenamefont {Sabooni}(2012)}]{Sabooni2012a}%
  \BibitemOpen
  \bibfield  {author} {\bibinfo {author} {\bibfnamefont {M.}~\bibnamefont
  {Sabooni}},\ }\href@noop {} {\bibfield  {journal} {\bibinfo  {journal} {et.
  al. (manuscript in preparation.)}\ } (\bibinfo {year} {2012})}\BibitemShut
  {NoStop}%
\bibitem [{\citenamefont {Nilsson}\ \emph {et~al.}(2004)\citenamefont
  {Nilsson}, \citenamefont {Rippe}, \citenamefont {Kr\"{o}ll}, \citenamefont
  {Klieber},\ and\ \citenamefont {Suter}}]{Nilsson2004}%
  \BibitemOpen
  \bibfield  {author} {\bibinfo {author} {\bibfnamefont {M.}~\bibnamefont
  {Nilsson}}, \bibinfo {author} {\bibfnamefont {L.}~\bibnamefont {Rippe}},
  \bibinfo {author} {\bibfnamefont {S.}~\bibnamefont {Kr\"{o}ll}}, \bibinfo
  {author} {\bibfnamefont {R.}~\bibnamefont {Klieber}}, \ and\ \bibinfo
  {author} {\bibfnamefont {D.}~\bibnamefont {Suter}},\ }\href@noop {}
  {\bibfield  {journal} {\bibinfo  {journal} {Physical Review B}\ }\textbf
  {\bibinfo {volume} {70}},\ \bibinfo {pages} {214116} (\bibinfo {year}
  {2004})},\ \bibinfo {note} {erratum: Physical Review B \textbf{71}, 149902(E)
  2005}\BibitemShut {NoStop}%
\bibitem [{\citenamefont {Timoney}\ \emph {et~al.}(2012)\citenamefont
  {Timoney}, \citenamefont {Lauritzen}, \citenamefont {Usmani}, \citenamefont
  {Afzelius},\ and\ \citenamefont {Gisin}}]{Timoney2012}%
  \BibitemOpen
  \bibfield  {author} {\bibinfo {author} {\bibfnamefont {N.}~\bibnamefont
  {Timoney}}, \bibinfo {author} {\bibfnamefont {B.}~\bibnamefont {Lauritzen}},
  \bibinfo {author} {\bibfnamefont {I.}~\bibnamefont {Usmani}}, \bibinfo
  {author} {\bibfnamefont {M.}~\bibnamefont {Afzelius}}, \ and\ \bibinfo
  {author} {\bibfnamefont {N.}~\bibnamefont {Gisin}},\ }\href@noop {}
  {\bibfield  {journal} {\bibinfo  {journal} {Journal of Physics B-atomic
  Molecular and Optical Physics}\ }\textbf {\bibinfo {volume} {45}},\ \bibinfo
  {pages} {124001} (\bibinfo {year} {2012})}\BibitemShut {NoStop}%
\bibitem [{\citenamefont {Nunn}\ \emph {et~al.}(2008)\citenamefont {Nunn},
  \citenamefont {Reim}, \citenamefont {Lee}, \citenamefont {Lorenz},
  \citenamefont {Sussman}, \citenamefont {Walmsley},\ and\ \citenamefont
  {Jaksch}}]{Nunn2008}%
  \BibitemOpen
  \bibfield  {author} {\bibinfo {author} {\bibfnamefont {J.}~\bibnamefont
  {Nunn}}, \bibinfo {author} {\bibfnamefont {K.}~\bibnamefont {Reim}}, \bibinfo
  {author} {\bibfnamefont {K.~C.}\ \bibnamefont {Lee}}, \bibinfo {author}
  {\bibfnamefont {V.~O.}\ \bibnamefont {Lorenz}}, \bibinfo {author}
  {\bibfnamefont {B.~J.}\ \bibnamefont {Sussman}}, \bibinfo {author}
  {\bibfnamefont {I.~A.}\ \bibnamefont {Walmsley}}, \ and\ \bibinfo {author}
  {\bibfnamefont {D.}~\bibnamefont {Jaksch}},\ }\href@noop {} {\bibfield
  {journal} {\bibinfo  {journal} {Physical Review Letters}\ }\textbf {\bibinfo
  {volume} {101}},\ \bibinfo {pages} {260502} (\bibinfo {year}
  {2008})}\BibitemShut {NoStop}%
\bibitem [{\citenamefont {Sabooni}\ \emph {et~al.}(2010)\citenamefont
  {Sabooni}, \citenamefont {Beaudoin}, \citenamefont {Walther}, \citenamefont
  {Lin}, \citenamefont {Amari}, \citenamefont {Huang},\ and\ \citenamefont
  {Kr\"{o}ll}}]{Sabooni2010}%
  \BibitemOpen
  \bibfield  {author} {\bibinfo {author} {\bibfnamefont {M.}~\bibnamefont
  {Sabooni}}, \bibinfo {author} {\bibfnamefont {F.}~\bibnamefont {Beaudoin}},
  \bibinfo {author} {\bibfnamefont {A.}~\bibnamefont {Walther}}, \bibinfo
  {author} {\bibfnamefont {N.}~\bibnamefont {Lin}}, \bibinfo {author}
  {\bibfnamefont {A.}~\bibnamefont {Amari}}, \bibinfo {author} {\bibfnamefont
  {M.}~\bibnamefont {Huang}}, \ and\ \bibinfo {author} {\bibfnamefont
  {S.}~\bibnamefont {Kr\"{o}ll}},\ }\href@noop {} {\bibfield  {journal}
  {\bibinfo  {journal} {Physical Review Letters}\ }\textbf {\bibinfo {volume}
  {105}},\ \bibinfo {pages} {060501} (\bibinfo {year} {2010})}\BibitemShut
  {NoStop}%
\end{thebibliography}%
\end{document}